\documentclass[conference]{IEEEtran}
\usepackage{amssymb}
\usepackage{amsmath}
\usepackage{txfonts}
\usepackage{graphicx}
\usepackage{multirow}
\usepackage{setspace}
\usepackage{hyperref}
\begin{document}
\bibliographystyle{IEEEtran}
\title{A Selection Region Based Routing Protocol for Random Mobile ad hoc Networks with Directional Antennas}
\author{\IEEEauthorblockN{Di Li, Changchuan Yin, and Changhai Chen}
\IEEEauthorblockA{Key Laboratory of Universal Wireless Communications, Ministry of Education\\
Beijing University of Posts and Telecommunications, China, 100876\\E-mail: dean.lidi@gmail.com,
ccyin@bupt.edu.cn, chenchanghai@gmail.com}}
\maketitle

\begin{abstract}
In this paper, we propose a selection region based multihop routing protocol with directional
antennas for wireless mobile ad hoc networks, where the selection region is defined by two
parameters: a reference distance and the beamwidth of the directional antenna. At each hop, we
choose the nearest node to the transmitter within the selection region as the next hop relay. By
maximizing the expected density of progress, we present an upper bound for the optimum reference
distance and derive the relationship between the optimum reference distance and the optimum
transmission probability. Compared with the results with routing strategy using omnidirectional
antennas in \cite{Di:Relay-Region}, we find interestingly that the optimum transmission probability
is a constant independent of the beamwidth, the expected density of progress with the new routing
strategy is increased significantly, and the computational complexity involved in the relay
selection is also greatly reduced. \footnote{This work is partially supported by the NSFC grants
60972073, 60971082, 60871042, and 60872049, the National Great Science Specific Project under
grants 2009ZX03003-011 and 2010ZX03001-003.}
\end{abstract}

%
\IEEEpeerreviewmaketitle

\section{Introduction}
Determining the capacity region of wireless networks has been an open problem for more than half a
decade. In the seminal work \cite{Capacity:Gupta}, Gupta and Kumar proved that the transport
capacity for wireless ad hoc networks, defined as the bit-meters pumped every second over a unit
area of the network, scales as $\Theta(\sqrt{n})$ in an arbitrary network, where $n$ is node
density. 
In \cite{Weber:Capacity}, Weber \emph{et al.} derived the upper and lower bounds on the
transmission capacity of spread-spectrum wireless ad hoc networks, where the transmission capacity
is defined as the product between the maximum density of successful transmissions and the
corresponding data rate, under a constraint on the outage probability. All these works address the
single hop transmission. Recently, Baccelli \emph{et al.} \cite{Baccelli:Aloha} proposed a spatial
reuse based multihop routing protocol, and derived the optimum transmission probability. In their
protocol, at each hop the transmitter selects the best relay to maximize the spatial density of
progress. By assuming each transmitter has a sufficient backlog of packets, Weber \emph{et al.} in
\cite{Weber:LongestEdge} proposed the longest-edge based routing protocol where each transmitter
selects a relay that makes the transmission edge longest. In \cite{Andrews:RandomAccess}, Andrews
\emph{et al.} defined the random access transport capacity with the maximum allowable total number
of transmissions per packet along multiple hops. In \cite{Di:Relay-Region}, Li\emph{ et al.}
proposed a selection region based multihop routing protocol to guarantee the message transmitted
towards the final destination, where the selection region is defined by two parameters: a selection
angle and a reference distance. By maximizing the expected density of progress, the author derived
an upper bound on the optimum reference distance, and the relationship between the optimum
reference distance and the optimum selection angle.

The above literatures only considered the wireless networks with omnidirectional antennas. In
\cite{Capacity-improvement-directional}, Yi \emph{et al.} investigated the wireless networks'
capacity using directional antennas, extending the network capacity with omnidirectional antennas
in \cite{Capacity:Gupta} to that with directional antennas. In \cite{Capacity-bounds-directional},
Spyropoulos \emph{et al.} discussed the network capacity gain one can achieve by using directional
antennas over that by using omnidirectional antennas and how these bounds are affected by important
antenna parameters like gain and beamwidth. In \cite{Multi-Channel-directional}, Dai \emph{et al.}
combined the multiple channels and directional antennas together and shown that they improve the
network capacity due to the increased network connectivity and reduced interference. However,
previous literature usually focuses on the scaling laws of the network capacity.

In this paper, we extend the former work in \cite{Di:Relay-Region} with omnidirectional antennas to
that with directional antennas, and derive the close-formed expected density of progress of the
network, which is defined as the number of packets progress toward their destinations in a unit
area of the network. Compared with the routing strategy using omnidirectional antennas in [1], due
to the directional antennas, the selection region based routing is implemented much easier and the
calculation burden for nodes to select the relay is also decreased.

The rest of the paper is organized as follows. Directional antenna model, network model and routing
strategy are described in Section II. The optimization for selection region and transmission
probability are presented in Section III. Numerical results and interpretations are given in
Section IV. Finally, Section V summarizes our conclusions.

\section{Network Model and Routing Protocol}
In this section, we first present the simplified directional antenna model, then define the network
model and the selection region based routing protocol using directional antennas.
\subsection{Directional Antenna Model}
In the study of wireless networks, the antenna model is often grouped into omnidirectional and
directional. Omnidirectional antenna radiates signals equally well in all directions, while
directional antenna has gain in the direction of the main lobe at which it is pointing. Thus, with
directional antennas the interference can be decreased and nodes located in each other¡¯s
neighborhood may transmit simultaneously, which increase spatial reuse of the channel. To simplify
the analysis, we model the power pattern of the directional antenna as a circular sector with angle
$\varphi$, where $\varphi$ is the beamwidth of the antenna, see Fig. 1. In the following analysis,
we assume the transmitters with directional antenna and the receivers with omnidirectional antenna,
which is called Directional Transmission and Omnidirectional Reception (DTOR) as mentioned in
\cite{Capacity-improvement-directional}.
\subsection{Network Model}
Assume nodes in the network follow a homogenous Poisson Point Process (PPP) with density $\lambda$,
and slotted ALOHA as the medium access control (MAC) protocol. During each time slot a node chooses
to transmit data with probability $p$, and to receive with probability $1-p$. Therefore, at a
certain time instant, the transmitters follow a homogeneous PPP ($\Pi_{Tx}$) with density
$p\lambda$, while the receivers follow another homogenous PPP ($\Pi_{Rx}$) with density
$(1-p)\lambda$. At each hop in the multihop transmissions, a transmitter tries to find a receiver
in $\Pi_{Rx}$ as relay. We consider the nodes are mobile, to eliminate the spatial correlation,
which is also discussed in \cite{Baccelli:Aloha}. We also assume that all transmitters use a fixed
transmission power $\rho$ and the wireless channel combines the large-scale path-loss and
small-scale Rayleigh fading. The normalized channel power gain is given by
\begin{equation}
G(d) = \frac{\gamma }{{{d^\alpha }}},
\end{equation}
where $\gamma$ denotes the small-scale fading, drawn from an exponential distribution of mean
$\frac{1}{\mu}$ with probability density function (PDF) $f_\gamma(x)=\mu \exp(-\mu x)$, and
$\alpha>2$ is the path-loss exponent.

For the transmission from transmitter $i$ to receiver $j$, an attempted transmission is successful
if the received signal-to-interference-plus-noise ratio (SINR) at the receiver $j$ is above a
threshold $\beta$. Thus the successful transmission probability over this hop with distance
$d_{ij}$ is given by
\begin{equation}
P_s=\Pr\left(\frac{\rho\gamma_0d_{ij}^{-\alpha}}{\sum_{k\in\Pi_{TX}\setminus\{i\}}\rho\gamma_i{d_{kj}}^{-\alpha}+\eta}>\beta\right),
\end{equation}
where $i\in \Pi_{Tx}$, $j\in \Pi_{Rx}$. Since we use directional antenna with beamwidth $\varphi$
for transmission, the interfers seen by a specific receiver follow a homogeneous PPP ($\Pi_{Tx'}$)
with density $p\frac{\varphi}{2\pi}\lambda$. Thus
$\sum_{k\in\Pi_{Tx'}\setminus\{i\}}\rho\gamma_i{d_{kj}}^{-\alpha}$ is the sum interference seen at
the receiver $j$, where $d_{kj}$ is the distance from interferer $k$ to receiver $j$, and $\eta$ is
the average power of ambient thermal noise. In the sequel we approximate $\eta=0$, which is
reasonable in interference-limited ad hoc networks. From \cite{Baccelli:Aloha}, the successful
transmission probability from transmitter $i$ to receiver $j$ is derived as
\begin{equation}
    P_s= \exp\left(p\frac{\varphi}{2\pi}\lambda t{d_{ij}}^{2}\right),
\end{equation}
where \[t = \frac{{2{\pi ^2}/\alpha }}{{\sin (2\pi /\alpha )}}{\beta ^{2/\alpha}}.\tag{3a}\]

\subsection{Routing Strategy with Directional Antennas}
%
Considering a typical multihop transmission scenario, where a data source sends information to its
final destination that is located far away, and it is impossible to complete this operation over a
single hop. Thus a multihop transmission is needed. In multihop wireless networks, if we assume
position-determined relays exist to ensure each hop shares the same distance that aggregates to
form the path from the data source to its final destination, the optimum transmission distance at
each hop is derived in \cite{Andrews:RandomAccess}. In this case, the transmission distance is used
to determine the location of a relay.
\begin{figure}
\begin{center}
\includegraphics[width=0.35\textwidth]{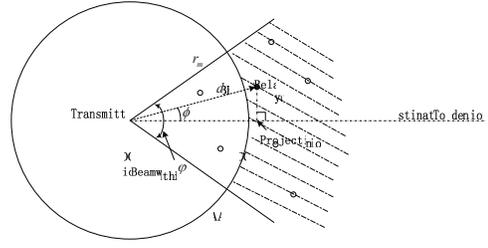}
\caption{Selection region with the directional antenna} \label{Scenario}
\end{center}
\end{figure}
In a practical case, nodes are usually randomly distributed, thus relays may not be located just
over the optimum transmission distance. To guarantee a relay existing at a proper position, we use
the selection region based multihop routing protocol with directional antennas. For each
transmitter along the route to the final destination, the selection region includes two parameters:
the beamwidth $\varphi$ and a reference distance $r_m$, as shown in Fig. 1, where the selection
region is defined as the region that is located within angle $\varphi$ and outside of the distance
$r_m$. Here, the transmitter is located in the circle center $O$, $\angle {BOC}=\angle
{AOC}=\varphi/2$, and $\overrightarrow{OC}$ points to the direction of the final destination. At
each hop, the transmitter selects the nearest receiver located in the selection region as the
relay.

Compared with the routing strategy using omnidirectional antennas in \cite{Di:Relay-Region}, the
selection region based routing with directional antennas can be implemented much easier. Since the
transmitters are equipped with directional antennas, only the receivers within the angle $\varphi$
can receive the radiated signals from the transmitters, thus the process to delimit the potential
receivers within the angle $\varphi$ do not need any calculation. However, for the routing protocol
with omnidirectional antennas, to determine if a receiver is located within the angle $\varphi$ or
not needs a complicated calculation, e.g., calculating the angle $\phi$ between the line from the
transmitter towards the final destination and the edge from the transmitter and the potential
receivers, and making comparisons between these $\phi$ with $\varphi$ to decide which nodes
are located within the selection region.
\section{Reference Distance and Transmission Probability Optimization}
%
In this section, we derive the optimum values of the transmission probability $p$ and the reference
distance $r_m$ for different beamwidth $\varphi$ by maximizing the expected density of progress.
\subsection{Upper Bound for Optimum Reference Distance}
As in \cite{Baccelli:Aloha}, the density of progress is defined as
\[D=p\lambda  \cdot P_{s} \cdot d \cos \phi,\]
where $P_s$ is the successful transmission probability defined in (2), $d \cos \phi$ is the
projection of the transmission distance $d$ along the line connecting the transmitter and the final
destination. Since the receivers follow a homogeneous PPP with density $\lambda(1-p)$, the
cumulative distribution function of the transmission distance $d$ is given as
\begin{equation}
    \Pr (d \leq r) = 1- \exp \left[ - \lambda (1 - p)\frac{\varphi }{2}({r^2} - r_m^2)\right],{r_m} \le r <
    \infty.
\end{equation}
Since $\phi$ is uniformly distributed over $[-\varphi/2,\varphi/2]$, which is independent of $d$,
the expected density of progress is given by
\begin{align}\label{E(d)}
 E[D] &= p\lambda^2 \int\limits_{r_m}^\infty \int\limits_{-\frac{\varphi}{2}}^{\frac{\varphi}{2}}  {{e^{ - p\frac{\varphi}{2\pi}\lambda t{x^2}}}} x \cos \phi {f_{d}}(x)d\phi dx\nonumber\\
  &= \lambda^2  p(1 - p)\Gamma \left(\frac{3}{2},kr_m^2\right){k^{ - 3/2}}\exp \left(\lambda (1 - p)\frac{\varphi }{2}r_m^2\right)\sin \left(\frac{\varphi
  }{2}\right),
\end{align}
where ${f_{d}}(x)$ is the probability density function of $d$ obtained from (4), $k =\frac{\lambda
\varphi}{2} (\frac{pt}{\pi} + (1 - p))$, $t$ is defined in (3a), and $\Gamma
\left(\frac{3}{2},kr_m^2\right) = \int\limits_{kr_m^2}^\infty {{e^{ - t}}{x^{\frac{3}{2} - 1}}dx} $
is the incomplete Gamma function.

To optimize the objective function in (5) with respect to the beamwidth $\varphi$, let us first
assume that $p$ is a constant, and try to derive the optimum value of $r_m$. For brevity, in the
following discussion, we write the objective function as $E$. Setting the derivative with respect
to $r_m$ as 0, after some calculations we have
\begin{align}\label{E/rm}
\frac{{dE}}{{dr_m^{}}} = &\exp \left(\lambda (1 - p)\frac{\varphi }{2}r_m^2\right)\nonumber\\
&\cdot\left[\Gamma \left(\frac{3}{2},kr_m^2\right)\lambda (1 - p)\varphi r_m^{} +
\frac{{d\Gamma\left(\frac{3}{2},kr_m^2\right)}}{{dr_m^{}}}\right] = 0,
\end{align}
where $\Gamma \left(\frac{3}{2},kr_m^2\right)$ is calculated as
\begin{equation}
\Gamma \left(\frac{3}{2},kr_m^2\right) = \Gamma \left(\frac{3}{2}\right) + \sqrt k {r_m}\exp \left(
- kr_m^2\right) - \frac{{\sqrt \pi  }}{2}erf\left(\sqrt k {r_m}\right).
\end{equation}
Therefore,
\begin{align}
 \frac{{d\Gamma\left(\frac{3}{2},kr_m^2\right)}}{{dr_m^{}}} & =  \sqrt k \exp \left( - kr_m^2\right)\left(1 - 2kr_m^2\right) - \sqrt k \exp \left( - kr_m^2\right) \nonumber\\
& = - 2{k^{3/2}}r_m^2\exp \left(- kr_m^2\right).
 \end{align}

Applying (8) to (6), we obtain
\begin{equation}
    \Gamma\left(\frac{3}{2},kr_m^2\right)\lambda \left(1 - p\right)\varphi {r_m} - 2{k^{3/2}}r_m^2\exp \left(- kr_m^2\right) = 0.
\end{equation}

Since it is difficult to analytically derive the exact solution for $r_m$ from (9), here we present
an upper bound for $r_m$. Since
\begin{align}\label{Gamma}
    \Gamma \left(\frac{3}{2},kr_m^2\right) & > \frac{1}{2}\left[\Gamma \left(1,kr_m^2\right) + \Gamma \left(2,kr_m^2\right)\right] \nonumber\\
    &= \frac{1}{2}\exp \left( - kr_m^2\right)\left(2 + kr_m^2\right),
\end{align}
using (10) in (9), we have
\begin{equation}
    \lambda (1 - p)k\varphi r_m^2 - 4{k^{3/2}}r_m^{} + \lambda (1 - p)\varphi  > 0.
\end{equation}
Therefore,
\begin{equation}
    {r_m} < \frac{{2{k^{3/2}} - \sqrt {4{k^3} - 2k{{[\lambda (1 - p)\varphi ]}^2}} }}{{k\lambda (1 - p)\varphi
    }}.
\end{equation}
\begin{figure}
\begin{center}
\includegraphics[width=0.4\textwidth]{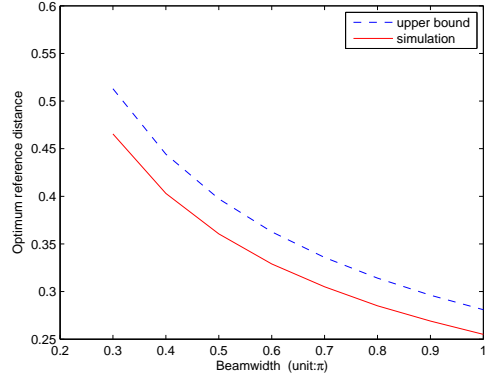}
\caption{Numerical results and the analytical upper bound for the optimum reference distance}
\label{2}
\end{center}
\end{figure}

In Fig. 2, we compare the upper bound of the optimum $r_m$ with the numerical results for different
beamwidth $\varphi$, when $p=0.1$.

\subsection{Jointly Optimizing the Reference Distance and the Transmission Probability}

Now, let us maximize the objective function by jointly optimizing $r_m$ and $p$, for different
beamwidth $\varphi$.

Rewrite (5) as
\[E =  \lambda^2  p(1 - p)\exp \left( - \frac{\varphi}{2\pi}\lambda ptr_m^2\right)\Gamma \left(\frac{3}{2},kr_m^2\right)\exp \left(
kr_m^2\right){k^{ - 3/2}}\sin \left(\frac{\varphi }{2}\right).\]

For brevity, we denote $\exp ( kr_m^2)$ as e and $\Gamma \left(\frac{3}{2},kr_m^2\right)$ as
$\Gamma$. With partial derivatives, we have
\begin{align}
\frac{{\partial E}}{{\partial r_m^{}}}=& \lambda^2  p(1 - p)\sin \left(\frac{\varphi
}{2}\right){k^{ -
3/2}}\exp \left( - \frac{\varphi}{2\pi}\lambda ptr_m^2\right)\nonumber\\
&\cdot \left[ -\frac{\varphi}{\pi}\lambda pt{r_m}\Gamma e + \Gamma \frac{{\partial e}}{{\partial
r_m^{}}} + e\frac{{\partial \Gamma }}{{\partial r_m^{}}}\right] = 0,
\end{align}
This holds only if
\begin{equation}
\Gamma \frac{{\partial e}}{{\partial r_m^{}}} + e\frac{{\partial \Gamma }}{{\partial
r_m^{}}}=\frac{\varphi}{\pi}\lambda pt{r_m}\Gamma e.
\end{equation}

 Since $k = \frac{\varphi \lambda(t-\pi)}{2\pi}p+\frac{\varphi\lambda}{2}$, there is $\frac{{\partial k}}{{\partial p}} =
\frac{\varphi\lambda(t-\pi)}{2\pi}$. To simplify things, we can then calculate the derivative with
respect to $k$ instead of $p$ as
\begin{flalign}
&\frac{{\partial E}}{{\partial k}} = \lambda^2  \sin \left(\frac{\varphi }{2}\right)\nonumber\\
&\cdot \left\{ \left[ -\frac{{ tr_m^2}}{{t - \pi }} - \frac{3}{2}{k^{ - 1}} + \frac{{1 - 2p}}{{p(1
- p)}}\frac{{2\pi }}{{\varphi \lambda(t - \pi )}}\right]\Gamma e + \left(\Gamma \frac{{\partial
e}}{{\partial k}} + e\frac{{\partial \Gamma }}{{\partial k}}\right)\right\}= 0.
\end{flalign}

By using the relationship $\frac{1}{2}r_m^{}\frac{{\partial e}}{{\partial r_m^{}}} =
k\frac{{\partial e}}{{\partial k}}$, $\frac{1}{2}r_m^{}\frac{{\partial \Gamma }}{{\partial r_m^{}}}
= k\frac{{\partial \Gamma }}{{\partial k}}$, and (14), we have
\begin{equation}
\Gamma \frac{{\partial e}}{{\partial k}} + e\frac{{\partial \Gamma }}{{\partial k}} =
\frac{{{r_m}}}{{2k}}\left(\Gamma \frac{{\partial e}}{{\partial {r_m}}} + e\frac{{\partial \Gamma
}}{{\partial {r_m}}}\right) = \frac{\varphi}{2\pi k}\lambda ptr_m^2\Gamma e.
\end{equation}

Applying (16) to (15), the following holds:
\begin{equation}
\left[ - \frac{{tr_m^2}}{{t - \pi }} - \frac{3}{2}{k^{ - 1}} + \frac{{1 - 2p}}{{p(1 -
p)}}\frac{{2\pi }}{{\varphi \lambda (t - \pi )}}\right]\Gamma e + \frac{\varphi }{{2k\pi }}\lambda
ptr_m^2\Gamma e = 0.
\end{equation}

After some calculation, we have
\begin{equation}
- \frac{{\varphi t}}{{2(t - \pi )}}\lambda r_m^2 - \frac{3}{2} + \frac{{[p + \pi /(t - \pi )](1 -
2p)}}{{p(1 - p)}} = 0.
\end{equation}

Given beamwidth $\varphi$, we use $\varphi$ and the optimum $p$ to express the optimum $r_m$ as
 \begin{equation}\label{Solution}
 {r_m} = \frac{1}{{\sqrt {\varphi \lambda } }}\sqrt {\frac{{2[p(t - \pi ) + \pi ](1 - 2p)}}{{p(1 - p)t}} - \frac{{3(t - \pi
 )}}{t}}.
 \end{equation}

Interestingly, by jointly optimizing $r_m$ and $p$, we find that the optimum transmission
probability $p$ in (19) is a constant independent of beamwidth $\varphi$, which is different from
the result we had in [1]. The proof is shown in Appendix.

Since in (19) $p$ is an constant, $r_m$ is only related to the beamwidth $\varphi$. Thus for a
given beamwidth $\varphi$, there is an associated optimum reference distance $r_m$ for the
transmitters to select next hop relay node, as shown in Fig. 4. Also note that in (19) $r_m$ scales
as $\lambda^{-1/2}$, which intuitively makes sense. This is because as the node density increases,
the interferers' relative distance to the receiver decreases as $\sqrt{\lambda}$, it requires a
shorter transmission distance by the same amount to keep the required SINR. By applying (19) in
(5), we observe that (5) becomes $N\sqrt{\lambda}$, where $N$ is a constant independent of
$\lambda$. This means that the maximum expected density of progress scales as
$\Theta(\sqrt{\lambda})$, which conforms to the results in \cite{Capacity:Gupta} and
\cite{Andrews:RandomAccess}.

\section{Numerical Results and Interpretations}
In this section, we present some numerical results based on the analysis in Section III. We choose
the path-loss exponent $\alpha$ as 3, the node density $\lambda$ as 1, and the outage threshold
$\beta$ as 10 dB. 

In Fig. 3, we show the optimum transmission probability obtained numerically vs. the beamwidth
$\varphi$. As shown in the figure, the optimum transmission probability $p$ is a constant, which
does not change with the beamwidth $\varphi$, this confirms our proof in Appendix. Thus, with the
selection region based routing with directional antennas, no matter how much the directional
antenna's beamwidth we select, the optimum transmission probability always keeps the same as
$p=0.12$. However, when we use the routing strategy with omnidirectional antenna, the optimum
transmission probability $p$ changes with the selection angle $\varphi$ (see Fig. 5 in [1]).
\begin{figure}
\begin{center}
\includegraphics[width=0.4\textwidth]{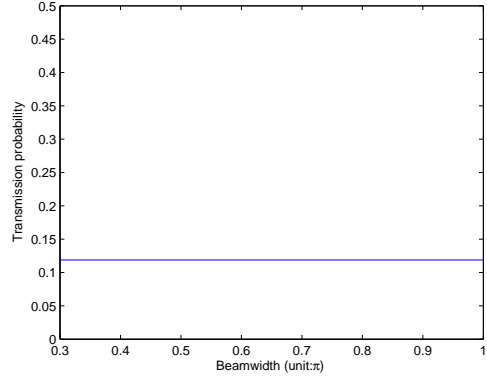}
\caption{The optimum transmission probability $p$ vs. the beamwidth $\varphi$ } \label{5}
\end{center}
\end{figure}
\begin{figure}
\begin{center}
\includegraphics[width=0.4\textwidth]{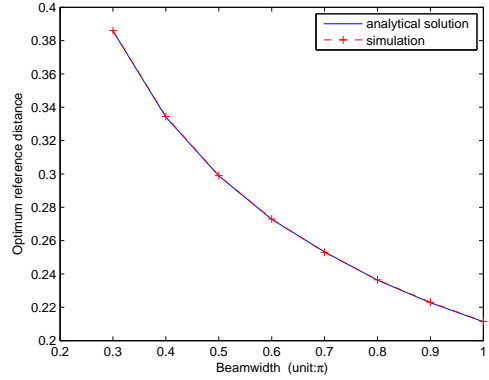}
\caption{The optimum reference distance vs. the beamwidth $\varphi$} \label{6}
\end{center}
\end{figure}

In Fig. 4, we compare the optimum reference distance $r_m$ obtained numerically with that derived
in (19), where $p$ is chosen optimally as the constant shown in Fig. 3. We see that increment of
the beamwidth $\varphi$ leads to the decease of the optimum reference distance. This can be
explained as follows: Increment of the beamwidth $\varphi$ means more interference seen by a
receiver; therefore, the optimum reference transmission distance should be decreased to guarantee
the quality of the received signal and the transmission successful probability.

\begin{figure}
\begin{center}
\includegraphics[width=0.4\textwidth]{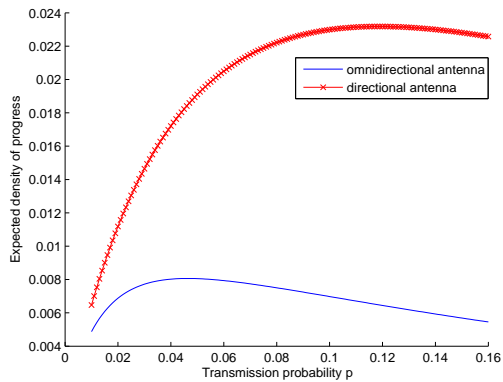}
\caption{Comparison of the expected density of progress for the routing protocol using directional
antennas with that using omnidirectional antennas} \label{7}
\end{center}
\end{figure}

In Fig. \ref{7}, we compare the expected density of progress of the routing protocol using
directional antennas with that using omnidirectional antennas in \cite{Di:Relay-Region}. From the
figure, we see that the expected density of progress for selection region based routing with
directional antennas owns a great advantage to that of the routing strategy with omnidirectional
antennas. This is because directional antennas can bring benefits such as reduced interference and
increased spatial reuse compared with omnidirectional antennas.

\section{Conclusions}
We propose a selection region based multihop routing protocol for wireless ad hoc networks with
directional antennas, where the selection region is defined by the beamwidth and the reference
distance. By maximizing the expected density of progress, we present some analytical results on how
to refine the selection region and the transmission probability. Compared with the routing strategy
with omnidirectional antennas in \cite{Di:Relay-Region}, the routing protocol in this paper owns
two main advantages: higher expected density of progress and less computational complexity when
choosing the next hop relay.


%
\appendix{Here, we prove that when jointly optimizing the reference distance $r_m$ and the transmission probability $p$,
the optimum transmission probability $p$ is a constant independent of beamwidth $\varphi$.}

Step 1: We consider the partial derivitive to the reference distance $r_m$, which is shown in (9),
here rewrite as
\begin{equation}
    \Gamma\left(\frac{3}{2},kr_m^2\right)\lambda \left(1 - p\right)\varphi {r_m} - 2{k^{3/2}}r_m^2\exp \left(- kr_m^2\right) =
    0,
\end{equation}
where $k =\frac{\lambda \varphi}{2} \left(\frac{pt}{\pi} + (1 - p)\right)$. Applying $k$ to (20),
we have
\begin{equation}
    \Gamma\left(\frac{3}{2},kr_m^2\right)\left(1 - p\right) - \left[\frac{pt}{\pi} + (1 - p)\right]\sqrt{{k}r_m^2}\exp \left(- kr_m^2\right) =
    0.
\end{equation}

Step 2: Now let us focus on the partial derivitive to the transmission probability $p$. As
mentioned in Section III-B, we calculate the derivitive with respect to $k$ instead, here rewrite
(15) as
\begin{flalign}
&\frac{{\partial E}}{{\partial k}} = \lambda^2  \sin (\frac{\varphi }{2})\nonumber\\
&\cdot \left\{ \left[ -\frac{{ tr_m^2}}{{t - \pi }} - \frac{3}{2}{k^{ - 1}} + \frac{{1 - 2p}}{{p(1
- p)}}\frac{{2\pi }}{{\varphi \lambda(t - \pi )}}\right]\Gamma e + \left(\Gamma \frac{{\partial
e}}{{\partial k}} + e\frac{{\partial \Gamma }}{{\partial k}}\right)\right\}= 0.
\end{flalign}

By using the same notation, defined in Section III-B, i.e., $e=\exp \left( kr_m^2\right)$ and
$\Gamma=\Gamma \left(\frac{3}{2},kr_m^2\right)$, we have
\begin{equation}
\frac{{\partial e}}{{\partial k}}=\exp ( kr_m^2)r_m^2.
\end{equation}
\begin{equation}
 \frac{{\partial \Gamma}}{{\partial k}}=-k^{\frac{1}{2}} r_m^3\exp ( kr_m^2).
\end{equation}

Applying (23) and (24) to (22), we have
\begin{flalign}
&\left[ -\frac{{ t k r_m^2}}{{t - \pi }} - \frac{3}{2} + \frac{{1 - 2p}}{{p(1 -
p)}}\frac{pt+\pi(1-p)}{{(t - \pi )}}\right]\Gamma e \nonumber \\
 &+ \Gamma \exp \left(kr_m^2\right)kr_m^2 - \left(k r_m^2\right)^{\frac{3}{2}}= 0.
\end{flalign}

Step 3: Applying (19) to (21) and (25), we see that in $kr_m^2$, $\varphi$ can be cancelled, thus
(21) and (25) become two equations that are independent of $\varphi$. Therefore, when jointly
optimizing $r_m$ and $p$, the optimum $p$ is a constant independent of the beamwidth $\varphi$,
only related to $t$ which is defined in (3a). And the numerical value for the optimum $p$ is given
in Fig. 3.



\end{document}